\documentclass[letters,useAMS,usenatbib]{mnras}

\usepackage{amsmath}
\usepackage{amssymb}
\usepackage{float}
\usepackage{graphicx}

\newcommand{\vect}[1]{\mathbfit{#1}}

\newcommand{\unit}[1]{\, \mathrm{#1}}
\newcommand{\msol}[0]{M_{\odot}}
\newcommand{\subscript}[1]{_{\mathrm{#1}}}
\newcommand{\fnbr}[2]{\, \mathrm{#1}\left({#2}\right)}

\newcommand{\sectionref}[1]{\hyperref[#1]{Section~\ref*{#1}}}

\newcommand{\columnfigure}[3]{\begin{figure}\includegraphics[width=\columnwidth]{#1}\caption{#2}\label{#3}\end{figure}}

\newcommand{\linefigure}[3]{\begin{figure*}\includegraphics[width=\linewidth]{#1}\caption{#2}\label{#3}\end{figure*}}

\newcommand{\scalelinefigure}[4]{\begin{figure*}\begin{center}\includegraphics[width=#1\linewidth]{#2}\caption{#3}\label{#4}\end{center}\end{figure*}}

\title[Small-scale dynamo - III. Cosmological simulations]{A small-scale dynamo in feedback-dominated galaxies - III. Cosmological simulations}
\author[M. Rieder, R. Teyssier]{Michael Rieder\thanks{Contact e-mail: \href{rieder@physik.uzh.ch}{rieder@physik.uzh.ch}} and Romain Teyssier\\
Institute for Computational Science\\
Centre for Theoretical Astrophysics and Cosmology\\ 
Universit\"at Z\"urich, 8057 Z\"urich, Switzerland}

\begin{document}

\maketitle

\begin{abstract}
Magnetic fields are widely observed in the Universe in virtually all astrophysical objects, from individual stars to entire galaxies, even in the intergalactic medium, but their specific genesis has long been debated.
Due to the development of more realistic models of galaxy formation, viable scenarios are emerging to explain cosmic magnetism, thanks to both deeper observations and more efficient and accurate computer simulations.  

We present here a new cosmological high-resolution zoom-in magnetohydrodynamic (MHD) simulation, using the adaptive mesh refinement (AMR) technique, of a dwarf galaxy with an initially weak and uniform magnetic seed field that is amplified by a small-scale dynamo driven by supernova-induced turbulence. As first structures form from the gravitational collapse of small density fluctuations, the frozen-in magnetic field separates from the cosmic expansion and grows through compression. In a second step, star formation sets in and establishes a strong galactic fountain, self-regulated by supernova explosions. Inside the galaxy, the interstellar medium becomes highly turbulent, dominated by strong supersonic shocks, as demonstrated by the spectral analysis of the gas kinetic energy. 
In this turbulent environment, the magnetic field is quickly amplified via a small-scale dynamo process and is finally carried out into the circumgalactic medium by a galactic wind.

This realistic cosmological simulation explains how initially weak magnetic seed fields can be amplified quickly in early, feedback-dominated galaxies, and predicts, as a consequence of the small scale dynamo process, that high-redshift magnetic fields are likely to be dominated by their small scale components. 
\end{abstract}

\begin{keywords}
early universe - galaxies: magnetic fields - methods: numerical - MHD - turbulence
\end{keywords}

\section{Introduction}

\linefigure{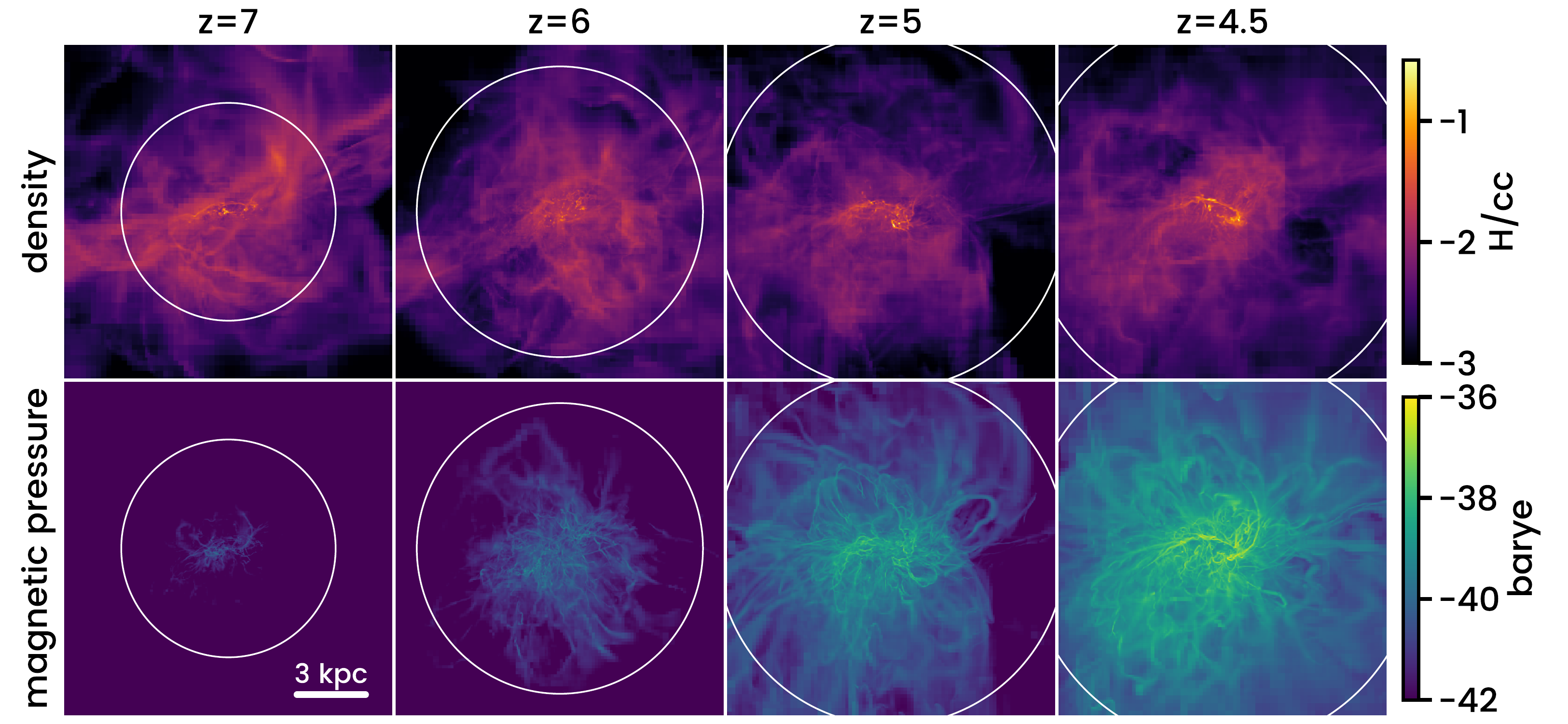}{Volume-averaged line-of-sight projections of gas density \emph{(top row)} and magnetic pressure \emph{(bottom row)} in the central 15 kpc around the galaxy at redshifts $z = 7,6,5,4.5$ (from \emph{left} to \emph{right}). The white circles mark the volume inside the virial radius.}{fig:images}

Magnetic fields are ubiquitous in the Universe. They are found in planets, stars, galaxies and may possibly permeate the intergalactic medium between them. Their origin could be primordial \citep{2013A&ARv..21...62D} or due to microphysical processes at later epochs, such as the Biermann battery \citep{M-1950ZNatA...5...65B} in shock fronts \citep{1997ApJ...480..481K} or ionization fronts \citep{2000ApJ...539..505G}, spontaneous fluctuations \citep{2012PhRvL.109z1101S} or fluctuations due to the Weibel instability \citep{2009ApJ...693.1133L} in the plasma of protogalaxies. Magnetic fields could also be released into the ISM by stars through stellar winds or supernova outbursts \citep{1973SvA....17..137B} or even by AGN jets \citep{2005LNP...664....1R} and subsequently diluted. 

Microphysical mechanisms like the Biermann battery can generate magnetic fields of the order of $10^{-20} \unit{G}$. The \cite{2016A&A...594A..19P} has set the upper limit for primordial magnetic fields (PMF) to $10^{-9} \unit{G}$, while \cite{2010Sci...328...73N,2011ApJ...733L..21D,2012ApJ...747L..14V} have set lower limits for the intergalactic field strength ranging from $10^{-18} \unit{G}$ to $10^{-15} \unit{G}$ based on $\gamma$-ray observations of blazar spectra. As far as galactic magnetic fields are concerned, observations from the Milky Way \citep{2009ApJ...702.1230T}, nearby galaxies \citep{2016A&ARv..24....4B} as well as high-redshift galaxies \citep{M-2008Natur.454..302B} reveal that they are stronger, around $10^{-6} \unit{G}$ and usually at equipartition with the turbulent energy density. \cite{2008ApJ...680..981R} detected field strengths up to $18 \unit{mG}$ in starburst galaxies but ordered galactic magnetic fields in the ISM are typically expected to be of the order of several $\unit{\mu G}$, similar in strength to the fluctuating components. \cite{2016ApJ...829..133K} found in intermediate redshift galaxies a clear correlation between large magnetic field strength and Mg II absorption, which indicates a link between strong outflows and a high magnetisation, as well as field as strong as a few $10^{-6} \unit{G}$ at epochs as early as $z \simeq 4-5$.

In order to explain this growth over several orders of magnitude in less than a Gyr, dynamo mechanisms are usually invoked which convert kinetic energy from gas flows into magnetic energy. Large-scale dynamos (LSD) are capable of amplifying magnetic fields coherently on large spatial scales, but on time scales that are too long to be consistent with the high-redshift observations. On the contrary, small-scale dynamos (SSD)  are very fast, with typical amplification time scales of the order of the smallest turbulent eddies turnover times  \citep{2012SSRv..169..123B}, 
which in high-redshift galaxies could be as fast as a few Myr \citep{2013A&A...560A..87S}.

Small-scale dynamos are very well studied both theoretically and experimentally. For the latter, laser driven experiments are currently being developed to study its development in a turbulent plasma \citep{2017PhPl...24d1404T}.
On the theoretical side, \cite{kazantsev1968} laid out the foundation of the SSD theory, for which \cite{M-1992ApJ...396..606K} found that the magnetic energy spectrum scales with the wavenumber as $k^{3/2}$ on scales larger than the resistive scale. It is generally admitted that SSD create fluctuating fields with a weak large-scale component and do not reach strict equipartition with the turbulent energy. 
This clearly does not conform with observations in nearby galaxies \citep{2015A&A...578A..93B}. It is likely that the magnetic fields we observe in present-day galaxies are not the result of just one single process, but probably a combination of various mechanisms.

Cosmological simulations performed with self-gravitating MHD codes have been reported since over a decade, first focusing on galaxy clusters and the intra-cluster medium 
\citep{2005JCAP...01..009D,2008A&A...482L..13D}, with very interesting results on the role  played by subsonic turbulence and moderate to high Mach number shocks in amplifying the field \citep{2014ApJ...782...21M,2015Natur.523...59M,2017MNRAS.464..210V}.
Later, zoom-in simulations of Milky Way-like galaxies have been performed with the SPH code GADGET \citep{2012MNRAS.422.2152B}, and with the moving-mesh code AREPO \citep{2014ApJ...783L..20P}.
They reported a fast amplification of the initial magnetic field up to saturation, invoking supersonic turbulence. 
They also observed that the redshift at which saturation is reached depends on the initial field strength, but the final field strength at saturation did not.

Most recently, \cite{2017MNRAS.469.3185P} reported similar results based on the Auriga suite of simulations, where small-scale dynamo amplification is observed in supersonic turbulence, 
until dynamo saturation is reached at 10\% of the turbulent energy level. The initial seed field  was chosen extremely high, of the order of $10^{-10} \unit{G}$, 
in order to observe the saturation at a high enough  redshift of $z = 2$.
All those experiments were using divergence cleaning methods for the magnetic field which suffers from possible problems with unphysical high magnetic field divergences.
The Constrained Transport (CT) technique is much more robust in this respect, and was previously only used in AMR codes, 
but has been recently adapted to Lagrangian moving-mesh codes in \cite{2016MNRAS.463..477M}.

With regard to realistic feedback mechanisms in cosmological MHD simulations, a more accurate treatment of feedback with cosmic rays (CR) physics was developed by \cite{2017MNRAS.465.4500P}, who included CR evolution equations in cosmological MHD simulations with AREPO. This opens new interesting mechanisms of magnetic field backreaction on the galaxy, such as a more strongly suppressed star formation in small galaxies due to the additional effect of CR pressure feedback. 

In \cite{2016MNRAS.457.1722R} (paper I) and \cite{2017arXiv170405845R} (paper II), we have studied how the turbulent environment in dwarf galaxies with strong feedback-driven winds is able to drive the SSD, amplifying even weak magnetic fields very rapidly, and how the resulting small-scale fields can be transformed into large-scale fields once feedback becomes weaker. Dwarf galaxies are the dominant galaxy population at high redshift, possibly responsible for cosmic re-ionisation \citep{2014ApJ...788..121K}. They are also the progenitors of the Milky Way satellites, which are useful laboratories to test our current galaxy formation paradigm.
In this letter, our intention is to build on our previous work and extend it to a more natural set-up with cosmologically realistic initial conditions, in order to make another step towards a better comprehension of the evolution of magnetic fields in the Universe. In \sectionref{sect:method}, we explain the numerical details of our simulation, the results of which are presented in \sectionref{sect:results}. We discuss these results in \sectionref{sect:discussion} and conclude with a future outlook in \sectionref{sect:conclusions}.

\section{Method}
\label{sect:method}

We used the Adaptive Mesh Refinement (AMR) code RAMSES \citep{2002AA...385..337T} to follow the cosmological evolution of a dwarf galaxy in a zoom-in simulation. This code simulates a self-gravitating magnetised plasma together with a collisionless fluid of dark matter and stars and additional physical sub-resolution processes such as gas cooling, star formation and supernova feedback. The ideal MHD equations are solved using a second order unsplit Godunov scheme \citep{2006JCoPh.218...44T} with a perfect gas equation of state. The gas is coupled to collisionless dark matter and stellar matter particles by the particle-mesh method. The solenoidal constraint
\begin{equation}\label{solenoidal} \nabla \cdot \vect B =0.\end{equation}
is implicitly fulfilled by the Constrained Transport (CT) method proposed to solve the induction equation by \citep{1966ITAP...14..302Y} and formulated by \cite{1988ApJ...332..659E}, thereby keeping the magnetic field divergence free. For a more detailed description of the numerical scheme, we refer the interested reader to \cite{2006A&A...457..371F}. Gas cooling is implemented using a standard H and He cooling function, with metal cooling (including both atomic  and fine-structure transitions). 

Star particles are created as a random Poisson process according to a Schmidt law as in \cite{2006A&A...445....1R} with an efficiency of $\epsilon_{*} = 1 \unit{\%}$. The effect of supernovae is modeled by releasing non-thermal energy into the ISM over a dissipation time scale of 20 Myr \citep{2013MNRAS.429.3068T} for $\eta_{\rm SN} = 10 \unit{\%}$ of the stars. These physics parameters for cooling, star formation and supernova efficiency have been selected and tested intensively in dwarf galaxy simulations with successful dynamo action in \cite{2016MNRAS.457.1722R} and \cite{2017arXiv170405845R} and were adopted here in a cosmological context. 

Our system of equations is formulated using  ``supercomoving variables'' to account for the expansion of the Universe in the Friedmann-Lema\^{i}tre-Robertson-Walker metric, as described by \cite{1998MNRAS.297..467M}. We chose to define the supercomoving magnetic field variable, using the scale factor $a$ as
\begin{equation}\widetilde{\vect{B}} = a^{5/2} \frac{\vect{B}}{B_*}\end{equation}
where $\vect B$ is the magnetic field in physical units and $B_* = \rho_*^{1/2} v_*$ is a fiducial scaling. Note that, although this definition is in contrast to the commonly used convention of $a^2$-scaling for $\widetilde{\vect{B}}$, both formulations are equivalent, and the induction equation in these supercomoving variables becomes
\begin{equation} \frac{\partial}{\partial \widetilde{t}} \; \widetilde{\vect{B}} = \widetilde{\nabla} \times \left( \widetilde{\vect{v}} \times \widetilde{\vect{B}} \right) + \frac{1}{2a} \frac{\mathrm{d}a}{d\widetilde{t}} \; \widetilde{\vect{B}}\end{equation}
thereby introducing only one new source term in the induction equation and conveniently leaving all the other MHD equations unchanged. 

We used the ``Multi-Scale-Initial-Conditions'' (MUSIC) toolkit developed by \cite{2011MNRAS.415.2101H}, together with the 2015 Planck cosmology parameters \citep{2016A&A...594A..13P} to generate our initial conditions. We ran a box with a comoving size of $7.5 \unit{Mpc}$ from redshift $z = 99$ until $z = 4.5$, zooming on a high resolution region around a dwarf-sized halo with $M = 1.75 \times 10^{10} \unit{\msol}$, selected from an initial unigrid dark matter only simulation. The mass resolution for dark matter particles in the zoomed region was $M = 1.5 \times 10^{4} \unit{\msol}$ and $M = 7.5 \times 10^{6} \unit{\msol}$ in the surrounding coarser region. We started with an initial effective resolution of $1024^3$ in the zoomed region. 

Further refinement levels are unlocked successively as the simulated universe expands and dark matter collapsed into haloes, ensuring that the physical resolution stays below $22.5 \unit{pc}$. While low-resolution cells are refined when the mass contained within exceeds 8 times a typical mass scale $M_*$, we decided to be more vigorous on refining the zoomed region by lowering that requirement for the finest resolution cells since dynamo amplification is very dependent on resolving the turbulent flow. The initial magnetic field was set to be uniform, aligned with the z axis and a field strength of $B_0 = 10^{-20} \unit{G}$ in physical units, giving it a conservatively low and therefore realistic starting value.

\section{Results}
\label{sect:results}

Images of volume-averaged line-of-sight projections of gas density and magnetic pressure are rendered in \autoref{fig:images} at various redshifts. As the galaxy evolves through time it grows substantially from mass inflows and
mergers, reaching a virial radius of $R_{200,c} = 8.6 \unit{kpc}$ at redshift $z = 4.5$. With densities becoming large enough to trigger star formation, feedback processes set in and drive turbulent winds which give rise to dynamo field amplification. Indeed, we can see the overall magnetic pressure rising with its filamentary structure typical for dynamo processes inside the galaxy and carried outside into the circumgalactic medium by winds.

\columnfigure{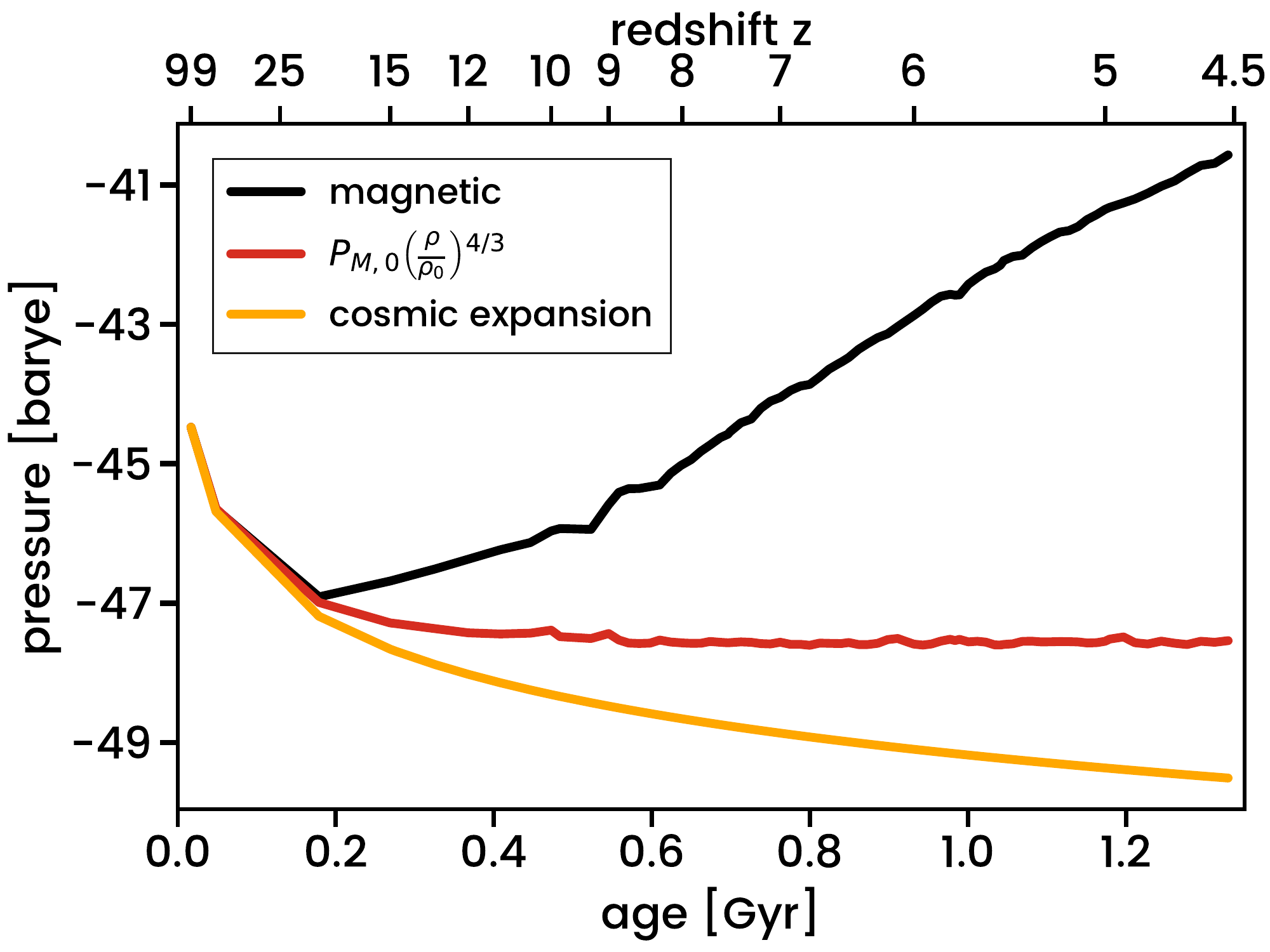}{Average magnetic pressure in physical, non-comoving units \emph{(black)} inside the zoomed region as a function of time and redshift. The yellow curve shows the $B \propto a^{-2}$ evolution for a trivially expanding universe and the red curve the expected scaling when the magnetic field follows the structure formation as $B \propto \rho^{2/3}$.}{fig:pmag}

\scalelinefigure{0.75}{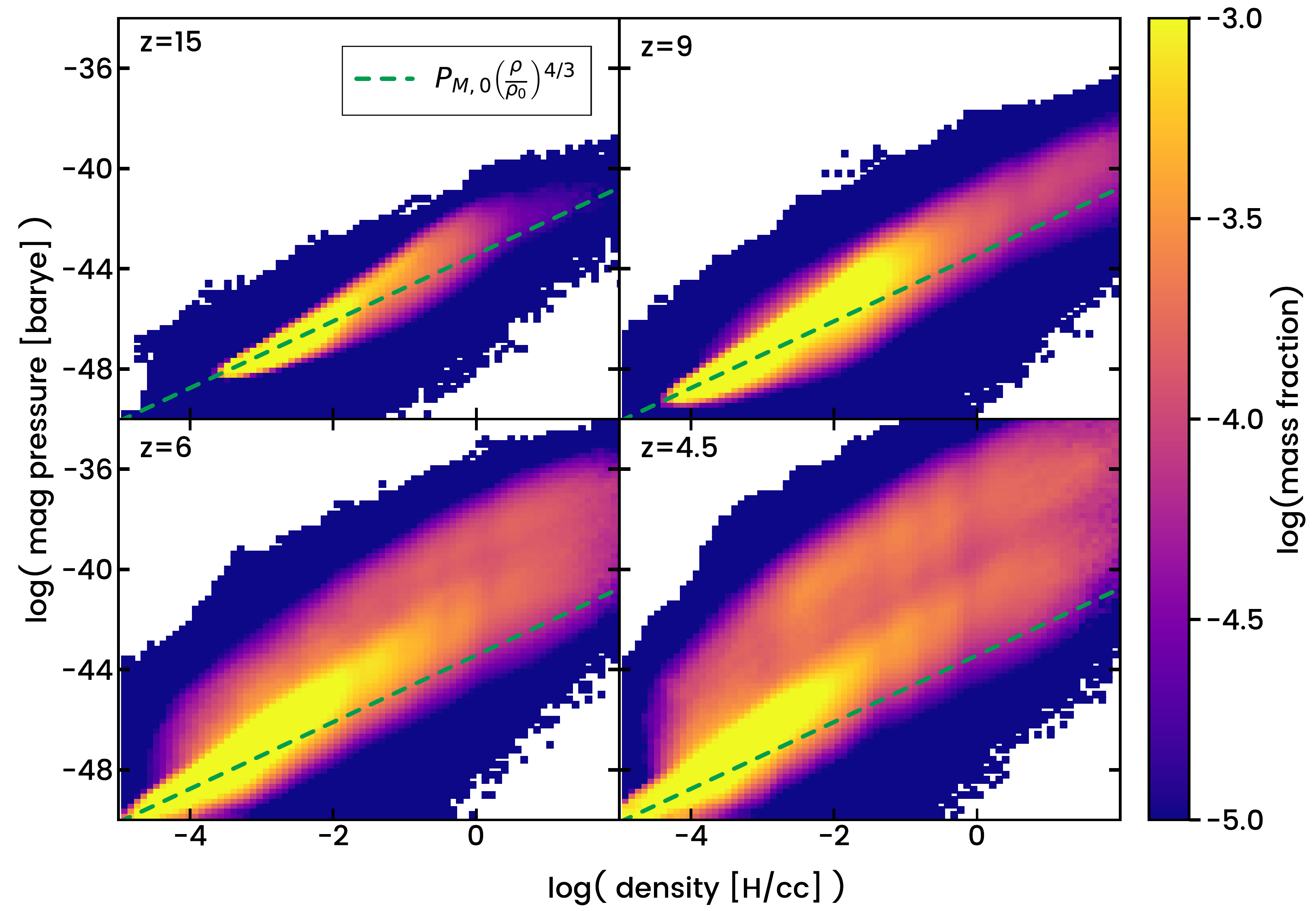}{Mass-weighted 2D histograms of magnetic pressure vs. gas density at redshifts $z=15,9,5,4.5$. The dashed green line indicates the magnetic pressure resulting from stretching or compression of the initial field when it follows the structure formation as $B \propto \rho^{2/3}$.}{fig:histo}

The evolution of the total magnetic pressure inside the zoomed region is plotted in \autoref{fig:pmag} together with the expected scaling of $B \propto \rho^{2/3}$ for frozen-in field lines as cosmic structures form and matter falls into dark matter halos. The initial evolution is dominated by cosmic expansion where gas density and magnetic field are both diluted as the universe expands and, subsequently, density fluctuations collapse into structures. We see this decrease until the first stars have formed and supernova explosions start to drive turbulent winds. This process sets off a continuous, self-regulating feedback process where steady mass infall triggers star formation and supernova winds push mass outwards. 

The resulting dynamo starts to amplify the field exponentially throughout its host galaxy's evolution history with an e-folding time $B \propto \fnbr{exp}
{t/\tau}$ of $\tau = 65 \unit{Myr}$, essentially undisturbed by merger events. This picture becomes clearer in \autoref{fig:histo}, where we plot mass-weighted 2D log-log histograms of gas density and magnetic pressure. Without any field amplification processes, magnetic pressure is expected to stay on the $P_M \propto \rho^{4/3}$ line tracking the stretching or compression of the initial magnetic field when it follows the structure formation. We can see that it starts to deviate from this line towards higher magnetic pressures at high densities where stars are forming and driving turbulent winds with their feedback processes. This process continues to higher and higher magnetic pressure which then also propagates to lower densities as magnetised winds transport magnetic energy from the dense central regions out to the circumgalactic medium.

\columnfigure{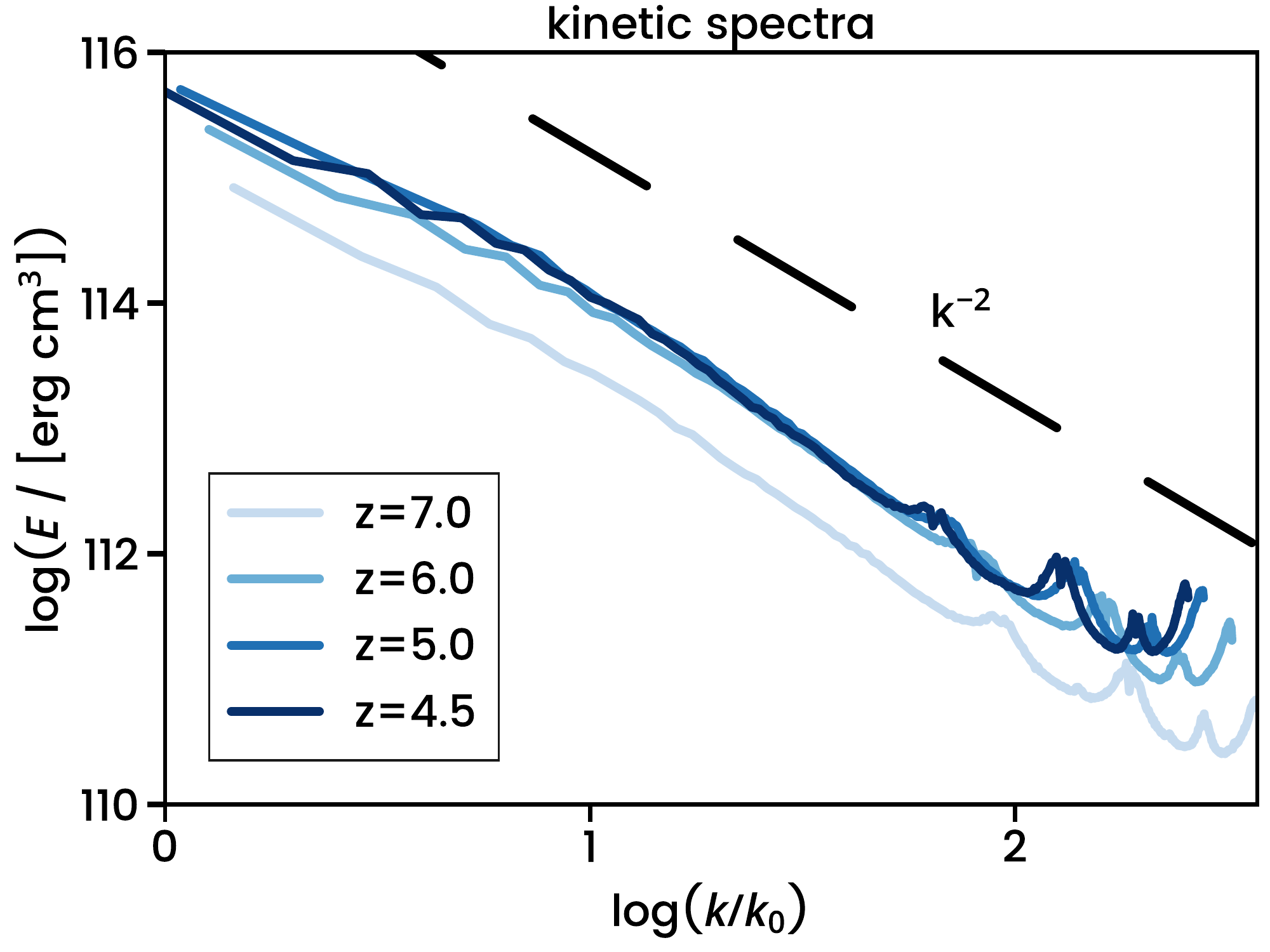}{Kinetic energy spectra in $512^3$ cubes centered on the halo at various redshifts, sampled at the grid resolution.}{fig:spectrakin}
\columnfigure{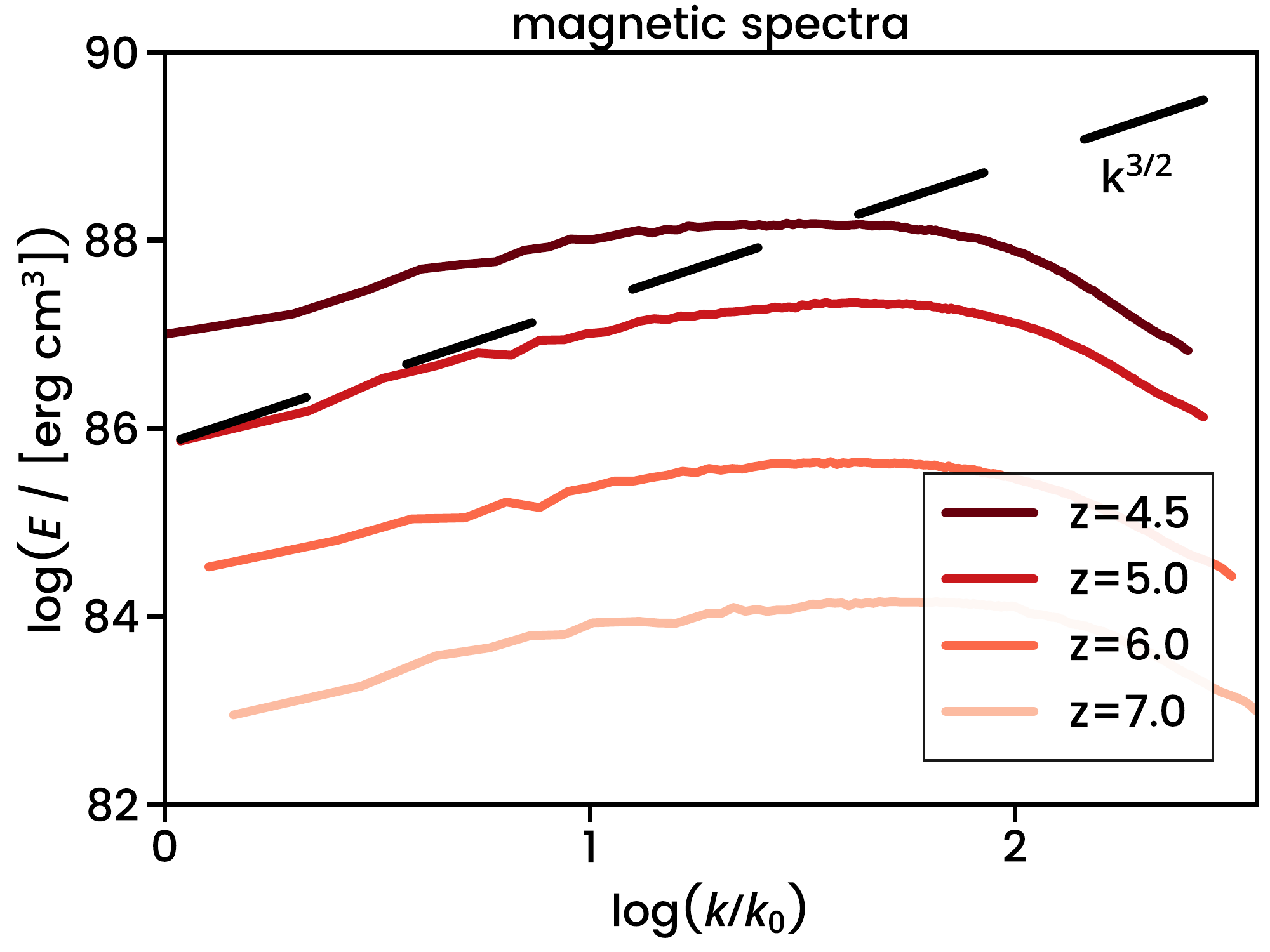}{Magnetic energy spectra in $512^3$ cubes centered on the halo at various redshifts, sampled at the grid resolution.}{fig:spectramag}

In \autoref{fig:spectrakin}, we plot the kinetic energy spectra at several redshifts. The spectra have a clear power-law shape $E\subscript{kin} \propto k^\alpha$ with a best fit for $\alpha$ ranging between $-2.04$ and $-2.13$. This slope is close to the theoretical value of $-2$ predicted for highly compressible, shock-dominated Burgers turbulence. Analogously, the magnetic energy spectra are also plotted for several redshifts in \autoref{fig:spectramag}. It develops the typical bottlenecked power-law shape with $E\subscript{mag} \propto k^{3/2}$ on larger scales as predicted by Kazantsev's theory, and falling off below the (numerical) resistive scale. 

We plot in \autoref{fig:velprofs} radial profiles of virial and tangential velocity, turbulent and sound speed, gas density, magnetic pressure and metallicity up to the virial radius, averaged in spherical shells around the centre of the galaxy at redshift $z=4.5$. The turbulent speed is comparable or higher than both the tangential velocity and the sound speed of the gas, leading to a supersonic flow with an average Mach number $\mathcal M \sim 2$. Both gas density and magnetic field strength fall off by two orders of magnitude from their peak value at the centre to the virial radius. The metallicity in comparison is much more uniform, as it is blown out by the galactic winds, with an average value of roughly $10 \unit{\%}$ solar metallicity.

\columnfigure{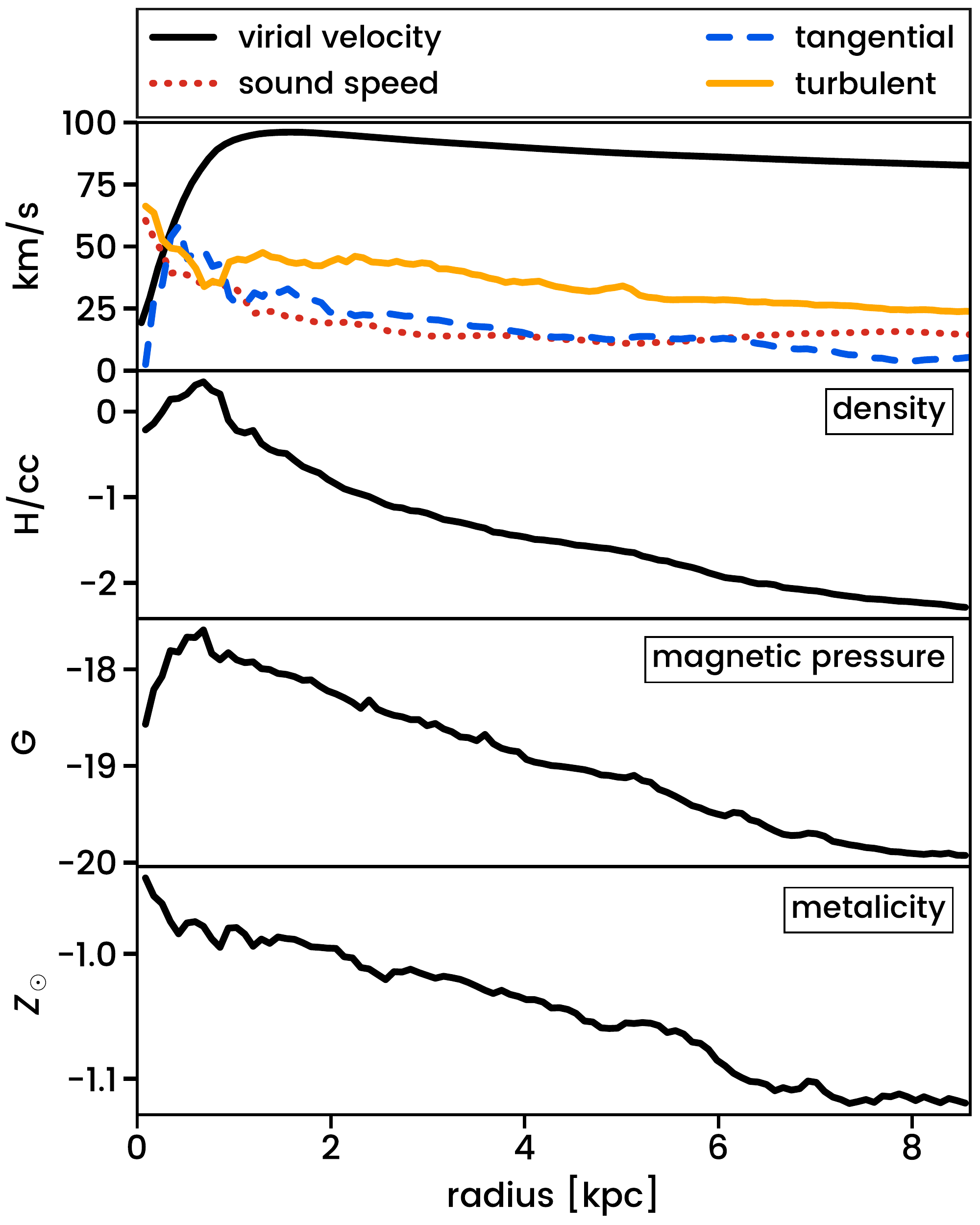}{Radial profiles of various characteristic speeds, gas density, magnetic pressure and metallicity averaged in spherical shells around the centre at redshift $z=4.5$.}{fig:velprofs}

\section{Discussion}
\label{sect:discussion}

We have performed MHD simulations of a zoom-in dwarf galaxy in a cosmological context, with an initial weak magnetic seed field, to study its evolution through cosmic time. Starting from an initially uniform universe with tiny density fluctuations and a spatially constant magnetic field, we see early structure formation where matter starts to fall into the potential wells of dark matter halos. This phase is characterised by dilution of the magnetic field as the universe expands and its scale factors increases. After redshift $z \sim 20$, this global effect is counter-balanced by the accelerated collapse of gas into haloes which also compresses magnetic field lines locally. 

As the first stars form inside these structures, they set the stage for a self-regulating mechanism of star formation and feedback-driven energy release. Galaxies form and this energy release gives rise to strong winds stirring the interstellar medium to become highly turbulent with kinetic energy spectra indicating a shock-dominated supersonic regime. This in turn leads to small-scale dynamo field amplification inside the galaxy and persistently rising magnetic field strengths, with magnetic energy spectra in conformance with Kazantsev dynamo theory. As magnetic pressure builds up in the central regions of the galaxy, it is carried out into the circumgalactic medium by magnetised winds. Even though the central galaxy is subject to several merger events throughout its cosmic evolution, they do not alter the turbulent flow process substantially and have no measurable effect on the dynamo mechanism.

These results are remarkable given the difficulty of capturing turbulence in galaxy simulations, even more so for a cosmological zoom-in, where the dynamic range of resolution from large scales down to the smallest possible in order to resolve the tiniest turbulent eddies is extraordinarily demanding in terms of computational efforts. The small-scale dynamo amplification rate essentially depends on the viscosity and magnetic diffusitivity of the medium, or the kinetic and magnetic Reynolds numbers respectively. As discussed in \cite{2017arXiv170405845R}, those characteristics are dominated by numerical resolution in this kind of simulation where any realistically attainable computational resolution is far from the required length scales of a realistic astrophysical plasma but can be extrapolated to real-world values, predicting full amplification from seed fields to saturation in just a few hundred Myr \citep{2010A&A...522A.115S,2012MNRAS.422.2152B}.

Therefore, a plausible scenario emerges where very weak seed fields, as predicted from Biermann battery mechanism,  are rapidly amplified inside galaxies by theSSD process to considerable strength at very high redshift just after the first stars have formed during very turbulent phases of a galaxy's history.
Our results confirm the findings of \cite{2017MNRAS.469.3185P}, where a small-scale dynamo was observed in Milky Way-like galaxies with comparable turbulent kinetic energy and Kazantsev magnetic energy spectra.

\section{Conclusions}
\label{sect:conclusions}

We have presented a cosmological simulation of a dwarf galaxy at  high redshift, where magnetic fields have been amplified exponentially from weak initial seed fields in a turbulent galactic environment.
They would certainly reach considerable strength, although slightly lower than equipartition, if we were not limited by our current computational capabilities. This work is an important step towards a comprehensive theory of magnetic field evolution from the early Universe after the Big Bang to the present time in a fully cosmological framework. Given the ever increasing computing power due to steady technological advance and improved software, it will become feasible to shed more light on this still open question. Future simulations will have to overcome the problem of the requested high resolution to resolve the correct growth rate for the dynamo mechanism and reduce the numerical resistivity. To this effect it is also worth considering the limits of the ideal MHD assumption, and how the varying ionisation fraction influences plasma properties, possibly leading to magnetic field diffusion and unsteady dynamo efficiency. Furthermore, there remains uncertainty on the nature of seed fields as there are numerous different viable mechanisms for their origin currently under discussion.

\section*{Acknowledgements}

This work was funded by the Swiss National Science Foundation SNF. All simulations were run on the Piz Dora cluster at the Swiss National Supercomputing Centre (CSCS) in Lugano, Switzerland.

\bibliography{michael,romain}

\end{document}